\documentclass[pra,aps,showpacs,lengthcheck]{revtex4-1}
\usepackage{amsmath}
\usepackage{graphicx}
\usepackage{epsfig}
\def\ket#1{|\,#1 \,\rangle}

\def\Lu{$^{175}\mathrm{Lu}^{+}$}

\def\um{\;\mu\mathrm{m}}

\def\GHz{\;\mathrm{GHz}}

\def\fref#1{Figure~\ref{#1}}
\def\tref#1{Table~\ref{#1}}

\def\S{^1\mathrm{S}_0}
\def\D{^3\mathrm{D}_1}

\def\Fp#1{F^\prime=\frac{#1}{2}}
\def\gs#1#2{\ket{\S,\frac{#1}{2},\pm\frac{#2}{2}}}
\def\es#1#2{\ket{\D,\frac{#1}{2},\pm\frac{#2}{2}}}
\def\gsalt#1#2{\ket{\S,\frac{#1}{2},\frac{#2}{2}}}
\def\esalt#1#2{\ket{\D,\frac{#1}{2},\frac{#2}{2}}}

\begin{document}
\date{\today}
\author{K. J. Arnold}
\email{cqtkja@nus.edu.sg}
\author{R. Kaewuam}
\author{A. Roy}
\author{E. Paez}
\author{S. Wang}
\author{M. D. Barrett}

\affiliation{  Centre for Quantum Technologies and Department of
 Physics, National University of Singapore, 3 Science Drive 2, 117543 Singapore}
\title{Observation of the $^1$S$_0$ to $^3$D$_1$ clock transition in $^{175}$Lu$^+$}
\begin{abstract}
We report the first laser spectroscopy of the $^1$S$_0$ to $^3$D$_1$ clock transition in $^{175}$Lu$^+$.  Clock operation is demonstrated on three pairs of Zeeman transitions, one pair from each hyperfine manifold of the $^3$D$_1$ term.  We measure the hyperfine intervals of the $^3$D$_1$ to 10 ppb uncertainty and infer the optical frequency averaged over the three hyperfine transitions to be $353.639~915~952~2\;(6)$ THz. The lifetime of the $^3$D$_1$ state is inferred to be $174^{+23}_{-32}$ hours from the M1 coupling strength.
\end{abstract}

\maketitle

\section{Introduction}

The development of optical atomic frequency standards has seen rapid progress in the past decade with evaluated fractional frequency uncertainties approaching the $10^{-18}$ level for both optical lattice clocks~\cite{SrYe,SrYe2,Ludlow} and single-ion clocks~\cite{AlIon,YbPeik}.   Recently the $^1$S$_0$ to $^3$D$_1$ transition in singly-ionised Lutetium has been identified as a promising clock candidate. The large fine and hyperfine structure splittings in combination with the technique of hyperfine averaging~\cite{MDB1} promises to realise an effective $J\;$=$\;0$ to $J^\prime\;$=$\;0$ transition with low sensitivity to magnetic fields. The differential scalar polarisability is expected to be sufficiently small for practical room temperature operation~\cite{kozlov2014optical,luprop2016}, and potentially negative, which would allow for micro-motion shifts to be eliminated~\cite{berkelandMicro,SrIon}. A negative polarisability in particular is critical for the realisation of a recent proposal for clock operation with large ion crystals~\cite{MDB2}. Such a realisation would provide a viable path to improve the stability of ion clocks to a level comparable with neutral lattice clocks.  

In this Article we report the first demonstration of direct excitation of the $^1$S$_0$ to $^3$D$_1$ M1 clock transition in $^{175}$Lu$^+$. For each of the three hyperfine transitions, we stabilise a probe laser to the average of a pair of Zeeman transitions to create effective $m_F=0\rightarrow m_{F^\prime}=0$ transitions. The frequency differences between the clock lasers give the hyperfine intervals independent of the linear Zeeman effect. Remaining residual frequency shifts due to second order Zeeman and ac-Stark effects are evaluated. This work represents the first steps towards establishing an optical clock based on $\mathrm{Lu}^+$ and demonstrates the simultaneous stabilisation of an optical oscillator to multiple hyperfine transitions separated by several GHz, a requirement for the future implementation of hyperfine averaging~\cite{MDB1}.  

\begin{figure}
\includegraphics[width=\columnwidth]{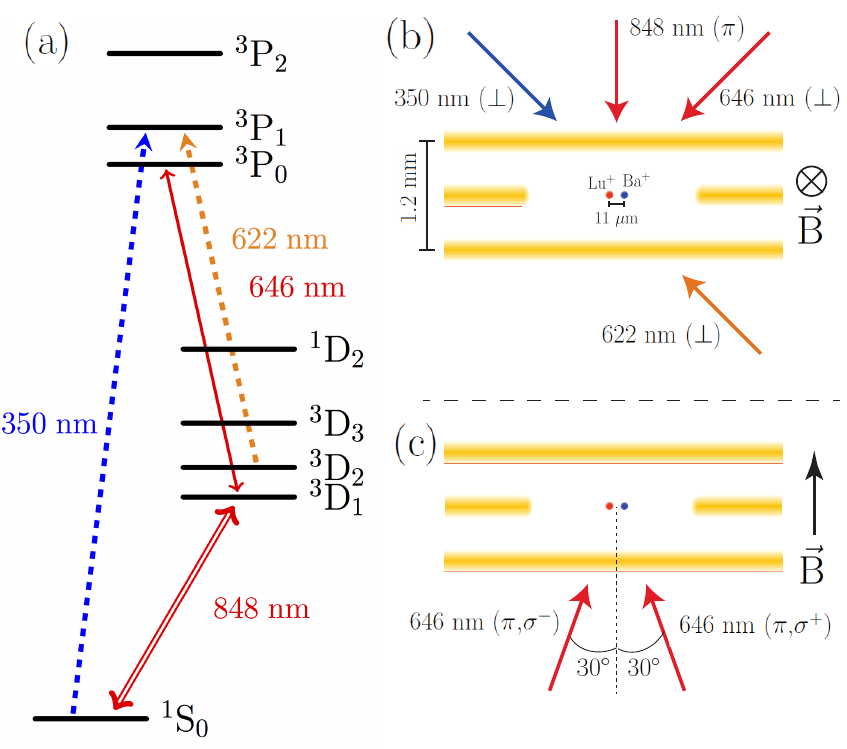}
\caption{[Color Online] (a) Level structure of Lu$^+$. The 350 nm and 622 nm transitions (dashed) provide optical pumping. The 646 nm transition (solid) provides detection and state preparation.  The 848 nm transition (double line) is the highly forbidden M1 clock transition. (b) and (c) Schematic representation of the Paul trap showing the geometry and polarisations of all addressing lasers. Doppler cooling lasers for Ba$^{+}$ (493 nm and 650 nm) are not shown.}
\label{setup}
\end{figure}

\section{Experimental Setup}
%\subsection{Ion Trap}
The experiments are performed in a four-rod linear Paul trap with axial end caps, shown schematically in \fref{setup}. The trap consists of four electropolished beryllium copper rods of 0.45 mm diameter arranged on a square of sides 1.2 mm in length. A 20.3 MHz rf potential is applied to two of the opposing diagonal rods via a helical quarter-wave resonator. A -0.1 V DC bias is applied to other two rods. Axial confinement is provided by 8 volts DC applied to the end caps, which are separated by 2 mm. In this configuration, the measured trap frequencies are $(\omega_x,\omega_y,\omega_z)/2\pi \approx (800,750,200)$ kHz. These frequencies are measured using a single $^{138}$Ba$^+$ ion which is present throughout to provide sympathetic cooling.

\begin{figure}
\includegraphics[width=\columnwidth]{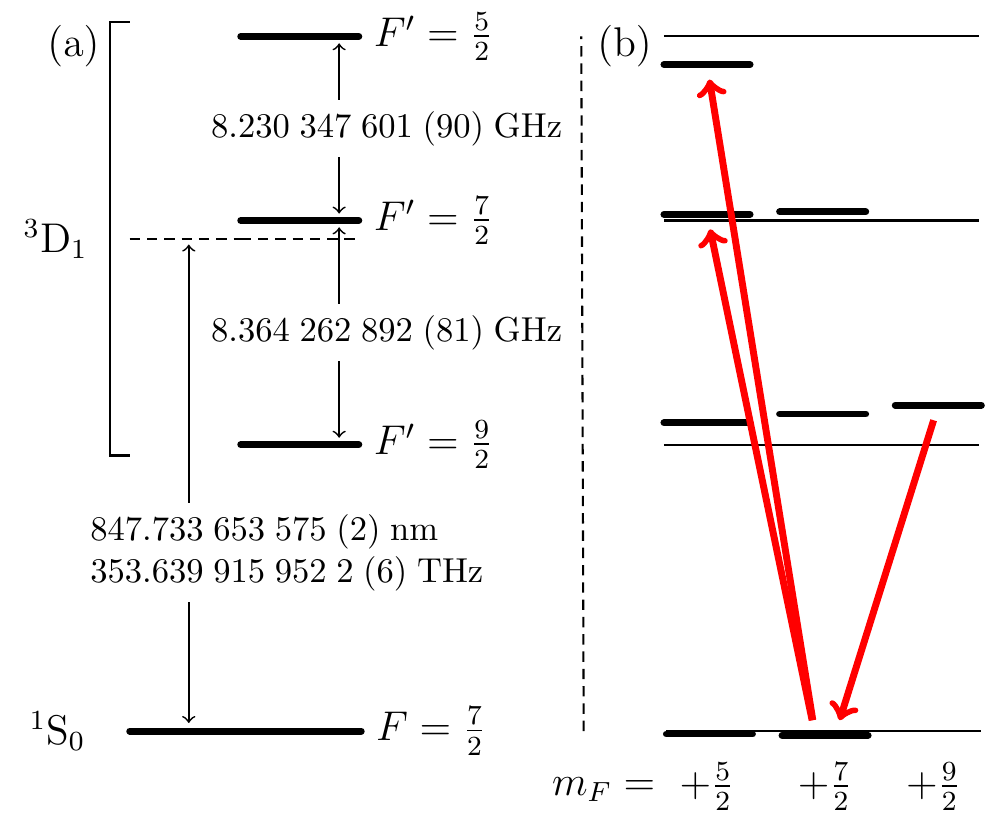}
\caption{[Color Online] The $^1$S$_0$ to $^3$D$_1$ clock transition in $^{175}$Lu$^+$. (a) Measured $\D$ hyperfine structure and average  optical transition frequency of the three hyperfine components. (b) Specific Zeeman transitions interrogated in the reported experiments. The three Zeeman transitions of opposite angular momentum to those shown are also interrogated. The thin solid lines represent the bare hyperfine energies in the absence of the Zeeman effect.}
\label{levelschematic}
\end{figure}

%lasers

%\subsection{Laser Systems}
The level structure of Lu$^+$ and the laser systems required are shown in \fref{setup}.  Lasers at 350 nm and 622 nm provide optical pumping to the $^3$D$_1$ state via the $^3$P$_1$ state.  Lasers at 646 nm couple to the nearly closed $^3$D$_1 \rightarrow \,^3$P$_0$ transition and are used for state detection of the $^3$D$_1$ state. These lasers propagate orthogonal to the applied magnetic field and are linearly polarised orthogonal to the magnetic field. Further details on optical pumping and detection in Lu$^+$ are provided in \cite{luprop2016}.  Detection is performed using a Bayesian scheme similar to that reported in~\cite{myerson2008high}. At the start of a detection cycle the probability of the ion being in the bright state is initialised to 0.5. Every 40 $\mu$s a field programmable gate array reevaluates this probability conditional on the number of photon arrivals in the previous time bin~\cite{luprop2016}. This continues until the probability that the ion is in the bright (dark) state reaches a preset threshold. For typical photon count rates of $\gtrsim 10$ counts/ms in the bright state, against a background of $\sim 1$ counts/ms, we achieve better than $99.9\%$ detection fidelity with mean detection time of $\approx 1.1$ ms.

State preparation is performed in $\D$ with an additional set of  646 nm lasers propagating at an angle of 30$^\circ$ with respect to a 0.4 mT applied magnetic field which defines the quantisation axis, see \fref{setup}(c). The polarisations are such that they consist of a linear combination of $\pi$ and either $\sigma^{\pm}$ in order to prepare the ion in either $\ket{\D,F^\prime\,$=$\,\frac{9}{2},m_{F^\prime}\,$=$\,\pm \frac{9}{2}}$ magnetic sub-state. We typically prepare $90\%$ of the population in either state.

%clock laser
The primary 848 nm clock laser is an interference-filter stabilised extended cavity diode laser of similar design to \cite{cateye}. The laser is locked to a 3.5 kHz linewidth high finesse ($F\approx400,000$) cavity with an ultra-low expansion glass spacer giving an expected laser linewidth of $\approx 1$ Hz ~\cite{SLS}. The primary laser is tuned near to the $\S$ to $\D(\Fp{9})$ transition. An additional frequency shift from an acousto-optic modulator (AOM) is used to rapidly switch between different Zeeman transitions within the $\D(\Fp{9})$ manifold. An auxiliary laser diode is optically injection locked to provide a stable optical power over the entire AOM frequency range required for interrogating the clock transitions.   A second laser for addressing either the $\Fp{7}$ or $\Fp{5}$ transition is phase locked to the primary laser with a frequency offset of either $\sim 8.4 \GHz$ or $\sim 16.6 \GHz$ as required. The second laser likewise utilises an AOM frequency shifter and injection locked auxiliary diode to provide stable intensity when addressing different Zeeman states.  For addressing the ion, both clock lasers propagate in the same spatial mode orthogonal to the magnetic field and have linear polarisation set parallel to the magnetic field. Selection rules for an M1 transition require $m_F - m_{F^\prime} = \pm 1$ in this configuration. We use the transitions illustrated in \fref{levelschematic} which satisfy this criteria.  

The experimental sequence for interrogation of the $\es{9}{9}\rightarrow  \gs{7}{7}$ transitions consists of the following procedure:
\begin{enumerate}
\item With both 622 nm and 350 nm pumping lasers and 646 nm detection lasers on, the Bayesian detection procedure is repeated until the ion is found in the bright $(\D)$ state\\
\item For 1 ms, Doppler cooling of $^{138}\mathrm{Ba}^+$ is applied to sympathetically cool Lu$^+$\\ 
\item Optical pumping by 646 nm light is applied for 400 $\mu$s to prepare in either $\es{9}{9}$ state \\
\item  Clock interrogation $\pi$ pulse on the $\es{9}{9} \rightarrow  \gs{7}{7}$ transition, typically 1 ms\\
\item Bayesian state detection procedure (mean time of 1.1 ms) \\
\end{enumerate}

\begin{figure}
\includegraphics{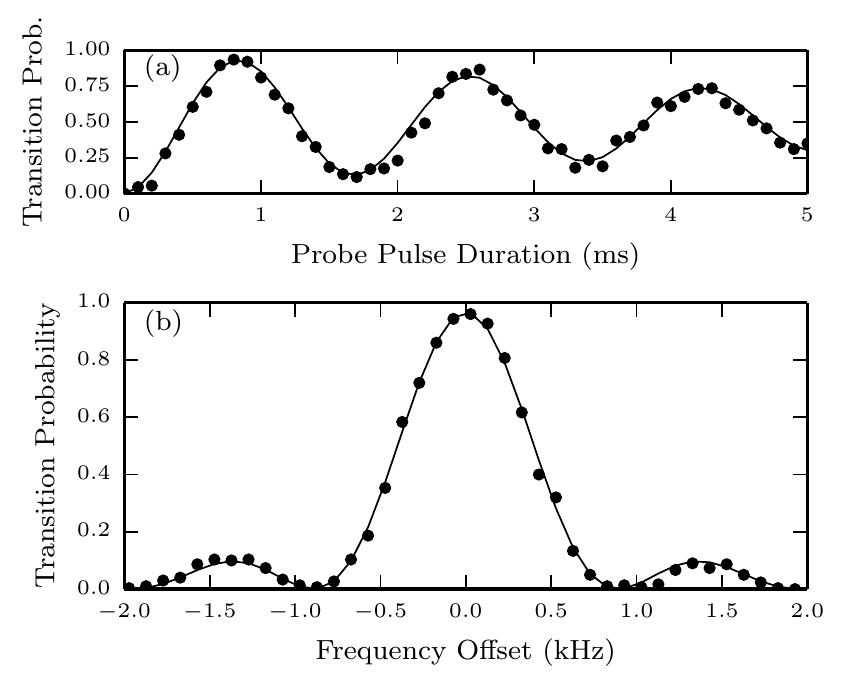}
\caption{Interrogation of the $\ket{\S,F\,$=$\,\frac{7}{2}, m_F \,$=$\,\frac{7}{2}} \rightarrow \ket{\D,F^\prime\,$=$\,\frac{7}{2},m_{F^\prime}\,$=$\,\frac{5}{2}}$ clock transition. Each point represents an average of 300 experiments. (a) Rabi flopping is observed by varying the interrogation time. The solid line fit is a sine function with exponentially decaying amplitude. (b) Transition probability when varying the laser frequency for fixed pulse time (1 ms). The fit is a fourier-limited Rabi lineshape with 98$\%$ contrast.}
\label{rabifig}
\end{figure}

 Since we are only able to optically pump into the $\es{9}{9}$ excited states, interrogation of clock transitions to other hyperfine transitions is most readily achieved using the $\es{9}{9} \rightarrow  \gs{7}{7}$ clock transition to conditionally transfer into the ground state. This is done by repeating steps 2-5 in the previous procedure until the ion is detected dark, which confirms transfer to the $\gs{7}{7}$ state. Next an interrogation pulse coupling to either the $\es{7}{5}$ or $\es{5}{5}$ state is then applied, followed by a final state detection step. \fref{rabifig} shows interrogation of the $\gsalt{7}{7}\rightarrow  \esalt{7}{5}$ transition when varying either the interrogation time or the probe laser frequency. 

To perform a precision measurement of the absolute transition frequency and hyperfine intervals, we must foremost eliminate the substantial linear Zeeman shifts, which are on the order of 7 MHz/mT for the most sensitive $\es{9}{9}$ states.  To do this we create effective $m_F=0$ to $m_{F^\prime}=0$ transitions by stabilising each laser to the average frequency of two Zeeman transitions with opposite angular momentum~\cite{bernard1998}. In practice this is achieved by alternately interrogating both Zeeman components at the half-width points of their line profile and separately tracking the mean and difference between the two transitions.

\section{Results}
\subsection{$^3$D$_1$ Hyperfine Spectroscopy}
\begin{figure}
\includegraphics{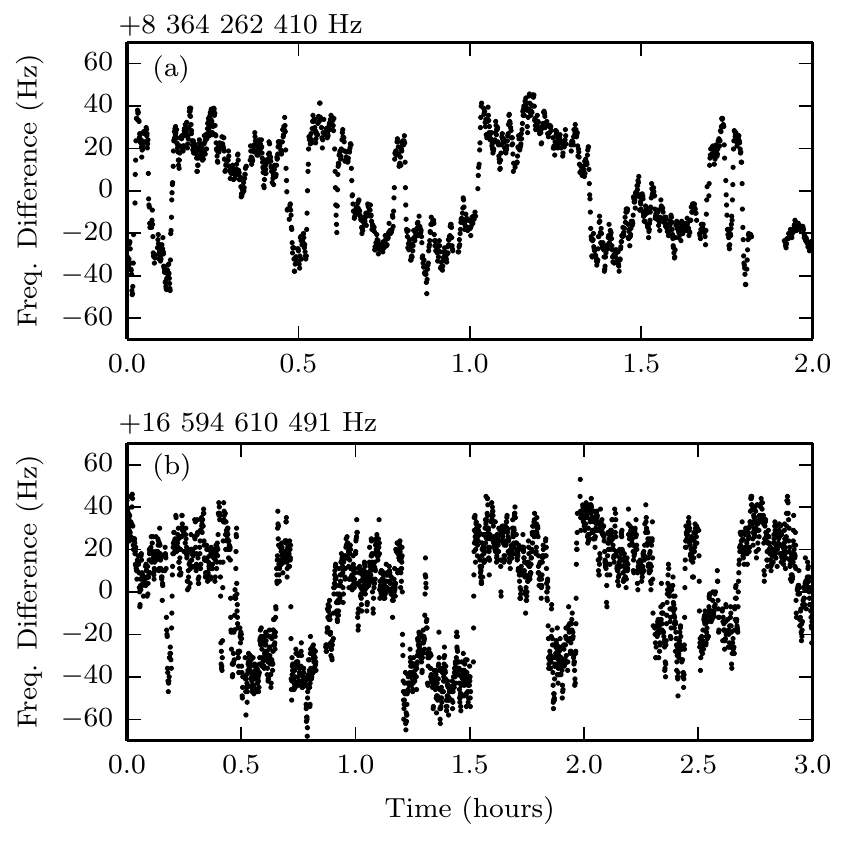}
\caption{Frequency difference between the primary clock laser stabilised to the $\es{9}{9} \rightarrow  \gs{7}{7}$ transition and the second clock laser which is stabilised to (a) the $\gs{7}{7} \rightarrow  \es{7}{5}$ transition or (b) the $\gs{7}{7} \rightarrow  \es{5}{5}$ transition. Discrete frequency jumps are due to a change in the ac-stark shifts when the two ions swap crystal positions (refer to text).}
\label{hyperfinefig}
\end{figure}

We measure the hyperfine interval between the $\Fp{7}$ and $\Fp{9}$ manifolds independent of linear Zeeman shifts by locking the primary clock laser to the average of the $\es{9}{9} \rightarrow  \gs{7}{7}$ transitions and the secondary laser to the average of the $\gs{7}{7} \rightarrow  \es{7}{5}$ transitions. We sequentially probe the half-width points of each of the four transitions for 50 experiments each to determine the frequency errors and then update all frequency offsets accordingly. Continuously correcting the frequency offsets for two hours, we measure an average frequency difference of $8~364~262~410$ Hz between the two lasers, see \fref{hyperfinefig}(a). The interval between the $\Fp{5}$ and $\Fp{9}$ manifolds is likewise measured by simultaneously locking the primary clock laser to the average of the  $\es{9}{9} \rightarrow  \gs{7}{7}$ transitions but instead stabilising the second laser to the average of the $\gs{7}{7} \rightarrow  \es{5}{5}$ transitions. The average frequency separation of the $\Fp{5}$ and $\Fp{9}$ lasers is measured for 3 hours to be $16~594~610~491$ Hz, see \fref{hyperfinefig}(b). Residual systematic frequency shifts must be accounted for to determine the hyperfine intervals.

The largest source of systematic uncertainty is the ac-stark shifts due to the 848 nm clock lasers. To evaluate these shifts, we first experimentally determine the dynamic differential scalar polarisability $\Delta \alpha_0$ and tensor polarisability $\alpha_2(\D)$ at the clock transition frequency. This is done by applying an auxiliary off-resonant 855 nm laser while the $F' = 9/2$ clock laser is locked to the atomic transition.  By comparing changes in the servo corrections under two different polarisations of the auxiliary laser, we can infer both $\Delta\alpha_0$ and $\alpha_2$ at $855\,\mathrm{nm}$.  We then extrapolate these values to $848\,\mathrm{nm}$, based on theoretical considerations, to obtain $\Delta \alpha_0=13.5\,(7)$ and $\alpha_2=-10.7\,(6)$ at the clock transition where both values are given in atomic units.  Details on the 855 nm polarisability measurements and extrapolation to 848 nm are provided in the Appendix~\ref{sec:dynamicpol}. We note that recent theoretical estimates~\cite{luprop2016} give $\Delta \alpha_0=16.5$ and $\alpha_2=-13$ at 848 nm, both $20\%$ higher than our experimental values.

During the hyperfine interval measurements, all clock lasers addressing the Lu$^+$ ion propagate in the same spatial mode which has a measured beam waist of 26 (1) $\mu$m. In \tref{errors}  we evaluate the ac stark shifts for the relevant transitions using the measured optical power for each of the respective clock lasers and the experimentally determined polarisabilities.  The stark shifts evaluated in \tref{errors} assume the beam is positioned exactly in between the two ions which are separated by 11 $\mu$m. In fact, when the Ba$^+$ and Lu$^+$ ions switch crystal positions there are discrete jumps in the ac-stark shift due to the difference in the clock laser intensity between the two positions, as can be clearly observed in the data shown in \fref{hyperfinefig}. From the temporal distribution of jumps we estimate that the Lu$^+$ ion spent an equal amount of time in each position to within $1\%$ and $3\%$ for datasets in  \fref{hyperfinefig}(a) and 3(b) respectively.  Additionally, from the magnitude of the jumps we estimate an axial displacement of the beam within the range of 1.5 to 3.5 um.  Consequently, when averaged over the ion positions, the AC stark shift decreases by at most $3\%$ over the range of estimated beam displacements.  Since this is considerably less than the uncertainty already present in the estimated AC stark shifts (\tref{errors}), the beam displacement is not taken into account.

The other significant systematic shift is due to the quadratic Zeeman effect. The independent servo of the linear Zeeman shift provides a continuous measurement of the magnetic field amplitude, limited only by our knowledge of the $g_{F^{\prime}}$ factors for the $\D$ level. We infer an average magnetic field of 0.400 (1) mT with rms stability of 21 nT from which we evaluate the residual quadratic Zeeman shifts in \tref{errors}.   Systematic shifts due to micromotion and quadrupole effects are evaluated to be $\sim$ 1 Hz or less and so are excluded.

Accounting for all the systematic shifts in \tref{errors}, we infer the $(\Fp{7}) - (\Fp{9})$ interval to be $8~364~262~892~(81)$ Hz and the $(\Fp{5}) - (\Fp{9})$ interval to be $16~594~610~494~(42)$ Hz. 

\subsection{$\S$ to $\D$ clock transition frequency}

We determine the optical frequency of the $\S$ to $\D$ transition by comparison of the primary clock laser to an optical frequency comb referenced to a GPS-disciplined Rubidium oscillator (Precision Test Systems GPS10RBN). From 20 hours of measurement with the clock laser stabilised to the effective linear field insensitive transition of the $F^\prime = \frac{9}{2}$ hyper-fine manifold, we infer the average optical frequency of the three hyperfine transitions to be $353.639~915~952~2\;(6)$ THz. The measurement stability and accuracy is completely limited by the GPS-disciplined microwave oscillator. 
 
\begin{table}[!t]
\centering
\begin{tabular}{ l@{\hskip 0.6cm}  r r@{\hskip 0.6cm} r r@{\hskip 0.4cm}  r r  }
\hline
$\D\ket{F^\prime,m_{F^\prime}}$&\multicolumn{2}{l}{$\ket{\frac{9}{2},\pm\frac{9}{2}}$} &\multicolumn{2}{l}{$\ket{\frac{7}{2},\pm\frac{5}{2}}$} &\multicolumn{2}{l}{$\ket{\frac{5}{2},\pm\frac{5}{2}}$} \\[.1cm]
\hline
ac stark (848 nm) &-62&(10)&-566&(91)&-326&(52)\\
quadratic Zeeman &$<$1 &  & 20.7 & (0.1) & 260.6 & (1.7) \\
%quadrupole  & & & & &  & \\
%second order doppler  & & & & &  & \\
\hline
Total Shift& -62&(10)& -545& (91)&-65&(52)\\
\end{tabular}
\caption{Systematic frequency shifts and uncertainties in Hz for transitions to the  $\D$ excited state listed in the first line of the table. In all cases the ground state of the respective transition is  $\ket{\S,F\,$=$\,\frac{7}{2}, m_F \,$=$\,\pm\frac{7}{2}}$. }
\label{errors}
\end{table}

\subsection{$\D$ lifetime}

The lifetime of the $\D$ state has been theoretically estimated to be 54 hours~\cite{luprop2016}. While this is too long to observe spontaneous decays directly, we can infer the lifetime from our observed M1 coupling. 
For a laser intensity at the ion of $3.53^{+14}_{-48}\times 10^{6}$ W/m$^2$ and $\pi$-polarisation, we observe a coupling rate of $\Omega = 323 (20)$ Hz on the $\es{9}{9} \rightarrow  \gs{7}{7}$ transition from which we infer the lifetime to be $174^{+23}_{-32}$ hours. 

\section{Conclusion}
In summary, this work represents the first steps towards establishing a Lu$^{+}$ optical clock by demonstrating spectroscopy on the $^1$S$_0$ to $^3$D$_1$ clock transition. This has been applied to measure the optical transition frequency and hyperfine structure of the $\D$ state for the \Lu isotope. The stability in the present experiment is principally limited by the use of magnetic field sensitive states. In turn, higher probe intensity was required to broaden the fourier limited linewidth to $\sim 1\;\mathrm{kHz}$ resulting in substantial ac stark shifts, on the order of $1\times10^{-12}$ fractional frequency shift. We note that the sensitivity to magnetic fields can be greatly suppressed by switching to the less abundant (2.6\%) isotope $^{176}\mathrm{Lu}^{+}$. This isotope has integer nuclear spin ($I\;$=$\;7$) for which first-order intensive $m_{F^\prime}=0$ states are available. With lower magnetic field sensitivity, it will be possible to extend the interrogation time to 100 ms or greater and consequently reduce the ac-stark shift to the order of $1\times10^{-16}$ or less. Further suppression can be achieved by hyper-Ramsey spectroscopy methods~\cite{hyperRamsey2010,hyperRamsey2012peik}. These straightforward improvements will greatly reduce the leading systematic shifts observed in the present work and it is reasonable to expect substantial improvements to both the stability and evaluated uncertainty in the near future. We would note that the fundamentally limited systematic shifts (Table I in~\cite{MDB2}) for the $^{176}\mathrm{Lu}^{+}$ isotope with implementation of the hyperfine averaging~\cite{MDB1} are expected to compare favourably with other leading ion clock candidates~\cite{YbPeik,AlIon} which currently achieve accuracies at the $10^{-18}$ level.

The long term prospects of the Lu$^+$ $^1$S$_0$ to $^3$D$_1$ transition as a clock candidate hinge on whether the differential static scalar polarisability $\Delta\alpha_0$ is negative and thus a viable candidate for realising a multi-ion clock~\cite{MDB2}. Recent theoretical estimates indicate that $\Delta\alpha_0$ is small, but with the sign indeterminate to within theoretical uncertainty~\cite{kozlov2014optical,luprop2016}. Even with the stability shown in this initial demonstration, experimental determination of the sign and magnitude of $\Delta\alpha_0$ may be possible by measuring and extrapolating from the dynamic polarisability at infrared wavelengths where intense laser sources are readily available. In particular, the Stark shifts induced by a 10.6$\um$ CO$_2$ laser would provide an unambiguous measurement of the sign and magnitude of $\Delta \alpha$ at dc. Experiments to perform these measurements are currently underway. 

\section{Acknowledgements}
This research is supported by the National Research Foundation, Prime Ministers Office, Singapore and the Ministry of Education, Singapore under the Research Centres of Excellence programme. It is also supported in part by A*STAR SERC 2015 Public Sector Research Funding (PSF) Grant (SERC Project No: 1521200080).

\appendix

\section{Dynamic Polarisability Measurement}
\label{sec:dynamicpol}

To determine the dynamic scalar polarisability $\Delta \alpha_0$ and tensor polarisability $\alpha_2$ at the clock transition, we measure the ac Stark shifts induced by an auxiliary 855 nm laser for both $\pi$ and $\perp$ polarisations. The wavelength 855 nm was used because it is the nearest wavelength to 848 for which we had a high power (600 mW single mode diode) source available. With the clock laser stabilised to the $\es{9}{9} \rightarrow \gs{7}{7}$ transition, we observe the change in the offset frequency of the clock relative to the reference cavity when we apply the 855 nm laser. The linear drift of the reference cavity, which was 0.104 (2) mHz/s at the time of this measurement, is subtracted from the servo corrections.  For the two polarisation settings of the 855 nm light, we measure the frequency shifts of clock transition to be $\delta_\pi = -241 \pm 8$ Hz and $\delta_\perp = -1945 \pm 6$ Hz. 

The waist of the 855 nm beam at the position of the ion is determined by measuring the ac-stark shift as a function of beam displacement. From a Gaussian fit to the measurements, we infer a waist of 42.6 (1.0) $\mu$m. The optical power deliver to the ion is 64.4 (1.3) mW with uncertainty determined by the specified accuracy of the optical power meter. From our implied optical intensity of $2.9~(0.2)\times10^5$ W/m$^2$, we determine the polarisabilities at 855 nm from the measured shifts to be:

\begin{align}
\Delta\alpha_0^{(855)} &= 13.0 \pm 0.7 \nonumber\\
\alpha_2^{(855)} &= -10.7 \pm 0.6 \nonumber
\end{align}
in atomic units.

We extrapolate to 848 nm using the matrix elements and polarisabilities reported in~\cite{luprop2016}. From the matrix elements, we estimate an 3.9\% increase in the differential scalar polarisability going from 855 to 848 nm. From the dynamic polarisabilities tabulated in Table IV of \cite{luprop2016}, we note the ratio $r\equiv \Delta \alpha_0/\alpha_2$ has a linear dependance on frequency from 1760 nm to 848 nm wavelengths. From this linear dependence, we estimate $r$ increases by 1.2\% going from 855 to 848 nm. We thus rescale the experimental polarisabilities to infer:
\begin{align}
\Delta\alpha_0^{(848)} &= 13.5 \pm 0.7 \nonumber\\
\alpha_2^{(848)} &= -11.0 \pm 0.6. \nonumber
\end{align}
We consider the extrapolation a systemic correction to our measured polarisabilities which does not introduce additional uncertainty comparable to the experimental uncertainties.

\bibliographystyle{h-physrev3}
%\bibliography{biblio}{}

%\pagebreak
%\includepdf{clocktransition_supp.pdf}

\end{document}